\documentclass[twocolumn,showpacs,amsmath,pre]{revtex4}

\usepackage{graphicx}
\usepackage{amsmath}
\usepackage{amssymb}
\usepackage{natbib}

\begin{document}

\title{The Contact Percolation Transition}

\author{Tianqi Shen$^{1}$}
\author{Corey S. O'Hern$^{2,1}$}
\author{M. D. Shattuck$^{3}$} 

\affiliation{$^{1}$Department of Physics, Yale University, New Haven,
  Connecticut 06520-8120, USA}

\affiliation{$^{2}$Department of Mechanical Engineering and Materials Science, Yale University, New
  Haven, Connecticut 06520-8286, USA}

\affiliation{$^{3}$Benjamin Levich Institute and Physics Department, The 
City College of the City University of New York, New York, New York 10031, USA}

\begin{abstract}
Typical quasistatic compression algorithms for generating jammed
packings of athermal, purely repulsive particles begin with dilute
configurations and then apply successive compressions with relaxation
of the elastic energy allowed between each compression step.  It is
well-known that during isotropic compression athermal systems with
purely repulsive interactions undergo a jamming transition at packing
fraction $\phi_J$ from an unjammed state with zero pressure to a
jammed, rigid state with nonzero pressure.  Using extensive computer
simulations, we show that a novel second-order-like transition, the
contact percolation transition, which signals the formation of a
system-spanning cluster of mutually contacting particles, occurs at
$\phi_P < \phi_J$, preceding the jamming transition.  By measuring the
number of non-floppy modes of the dynamical matrix, and the
displacement field and time-dependent pressure following compression,
we find that the contact percolation transition also heralds the onset
of complex spatiotemporal response to applied stress.  Thus, highly
heterogeneous, cooperative, and non-affine particle motion occurs in
unjammed systems significantly below the jamming transition for
$\phi_P < \phi < \phi_J$, not only for jammed systems with $\phi >
\phi_J$.

\end{abstract}

\pacs{
83.80.Fg,
64.60.ah, 
61.43.-j,
61.43.Gt
}

\maketitle

\begin{figure*}
\includegraphics[width=0.9\textwidth]{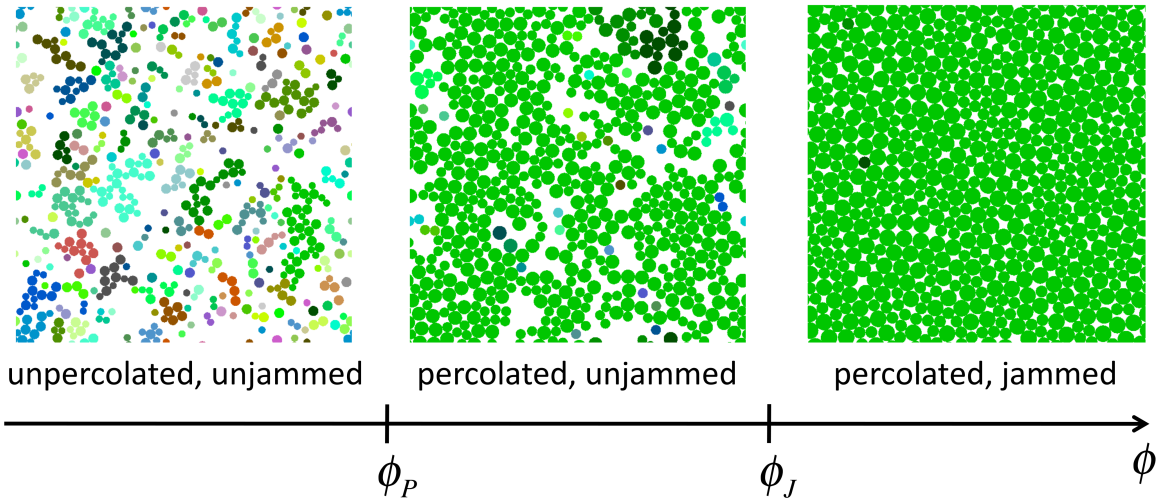}
\vspace{-0.1in}
\caption{(Color online) Typical snapshots from the quasistatic
isotropic compression algorithm to generate static particle
configurations as a function of packing fraction $\phi$. Particles
with a given color shading belong to the same cluster of mutually
contacting particles.  For (left) $\phi < \phi_P$, the system is
unjammed, and the largest cluster of contacting particles does not
percolate.  The largest cluster begins to percolate for (middle)
$\phi_P < \phi < \phi_J$, but the system remains unjammed since it
possesses nontrivial floppy modes.  For (right) $\phi > \phi_J$, the
percolating cluster of contacting particles has become rigid except
for a small number of rattlers~\cite{footnote}, and the system is
jammed.}
\label{fig1}
\vspace{-0.1in}
\end{figure*}

The jamming transition in athermal, purely repulsive particulate
systems, such as granular media, foams, and colloids, has been
characterized extensively in computer simulations~\cite{j} and
experiments~\cite{behringer}.  For example, when model frictionless
spheres are compressed to packing fractions above $\phi_J$, static
particle configurations transition from unjammed to jammed with
nonzero pressure (and elastic energy) and a rigid backbone of
contacting particles with all positive nontrivial
eigenvalues~\cite{barrat} of the dynamical matrix. Signatures of the
jamming transition, such as anomalous scaling of the zero-frequency
shear modulus with packing fraction~\cite{vanhecke} and diverging
length scales~\cite{cnls,silbert} associated with cooperative particle
rearrangements, have been investigated in thermal systems~\cite{zhang}
in the zero-temperature limit and in sheared
systems~\cite{tighe,heussinger1,heussinger2,vagberg} in the zero-shear
rate limit, but mainly for packing fractions near and above $\phi_J$.

However, there have been few detailed studies of the structural and
mechanical properties of {\it unjammed} athermal particulate systems
well below $\phi_J$. As shown in Fig.~\ref{fig1}, typical quasistatic
compression algorithms used to generate static packings in experiments
start with a dilute collection of particles, and the sample is
successively compressed by small amounts with energy relaxation allowed
between each compression step.  For $\phi < \phi_J$, the
configurations are not completely rigid, and thus after each small
compression, particles can rearrange until all interparticle forces
are zero.  Despite this, the response of unjammed packings to
compression and other perturbations can be highly heterogeneous,
cooperative, and non-affine.  Thus, an important, unanswered question
is at what packing fraction does complex spatiotemporal response
first occur in unjammed particulate systems?

In this letter, we describe computational studies of a novel
second-order-like transition---the contact percolation transition at
$\phi_P$---in athermal particulate systems of purely repulsive,
frictionless disks that signals the formation of a system-spanning
cluster of connected non-force-bearing interparticle contacts and the
onset of complex spatiotemporal response well below the jamming
transition.  These systems display robust scaling behavior near
$\phi_P$, but with a correlation length exponent that differs
from the corresponding values for random
continuum~\cite{gawlinski} and rigidity percolation~\cite{thorpe}.  In
addition, we find that the the number of `blocked' degrees of freedom,
and the particle displacement and stress relaxation time in
response to compression begin to increase significantly for $\phi >
\phi_P$.  These results emphasize that highly cooperative and glassy
dynamics occur in compressed athermal systems significantly below the jamming
transition, not only for jammed systems with $\phi > \phi_J$.
   
We focus on systems composed of bidisperse frictionless disks that 
interact via the purely repulsive linear spring potential: 
\begin{equation}
\label{pot}
V(r_{ij}) = \frac{\epsilon}{2} \left( 1 - \frac{r_{ij}}{\sigma_{ij}} \right)^2 \theta\left(1-\frac{r_{ij}}{\sigma_{ij}}\right),
\end{equation} 
where $\epsilon$ is the characteristic energy scale, $\theta(x)$ is 
the Heaviside step function, $r_{ij}$ is the
separation between the centers of disks $i$ and $j$, and $\sigma_{ij}
= (\sigma_i+\sigma_j)/2$ is their average diameter.  We chose a
$50-50$ (by number) mixture of large and small disks with diameter
ratio $\sigma_l/\sigma_s=1.4$ to inhibit crystallization~\cite{j} and
implemented periodic boundary conditions in a unit square.  We
employed a quasistatic isotropic compression algorithm to generate
static packings over a range of packing fractions~\cite{gao}. We
initialize each system with random particle positions at $\phi=0$ and
zero velocities.  We then compress the system in steps of $\Delta \phi
= 10^{-3}$ and relax the small particle overlaps after each step by
solving Newton's equations of motion in the overdamped limit,
\begin{equation}
\label{newton}
m {\vec a}_i = \sum_j {\vec F}(r_{ij}) - b {\vec v}_i,
\end{equation}
where $m$ and ${\vec a}_i$ are the particle mass and acceleration,
${\vec F}(r_{ij}) = -dV(r_{ij})/dr_{ij} {\hat r}_{ij}$, ${\hat
r}_{ij}$ is the unit vector connecting the centers of particles $i$
and $j$, and ${\widetilde b} = b \sigma_s/\sqrt{m \epsilon}$ is the
damping coefficient, until the total potential energy per particle falls
below a specified tolerance $V/\epsilon N < V_{\rm tol} = 10^{-16}$.
We studied two values for the damping coefficient, ${\widetilde b} =
1$ and ${\widetilde b} \rightarrow \infty$ using steepest descent
dynamics.  We continue compression steps followed by relaxation until
the systems jam at a configuration dependent $\phi_J \approx 0.84$.

\begin{figure}
\includegraphics[width=1.05\columnwidth]{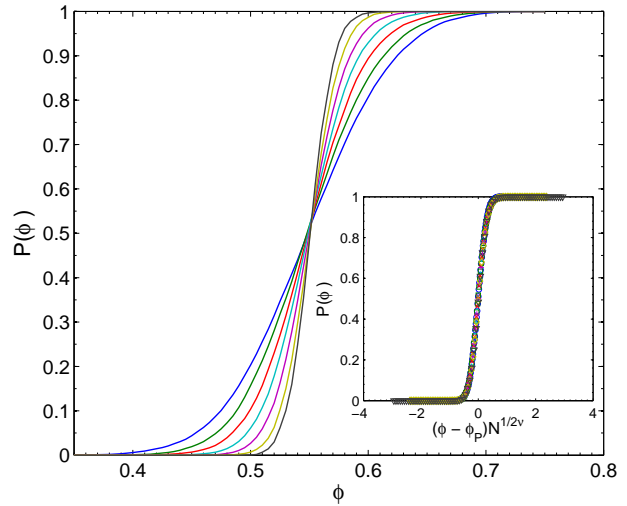}
\vspace{-0.3in}
\caption{Percolation probability $P(\phi)$ that the system possesses a
system-spanning cluster (in either the $x$- or $y$-direction)
immediately following a compression step $\Delta \phi=10^{-3}$ versus
packing fraction $\phi$ for $N=100$, $200$, $400$, $800$, $1600$,
$3200$, and $6400$ particles (from left to right) averaged over $400$
configurations. Inset: Same as the main figure except the horizontal
axis is scaled by $(\phi-\phi_P) N^{1/2\nu}$, where $\phi_P = 0.549
\pm 0.001$ and $\nu =1.70 \pm 0.09$.}
\vspace{-0.1in}
\label{fig2}
\end{figure}

In Fig.~\ref{fig2}, we characterize the contact percolation transition
by plotting the probability $P(\phi)$ that the system forms a
system-spanning network of interparticle contacts in either the $x$-
or $y$-direction, where contact is determined by $r_{ij} \le
\sigma_{ij}$, at each $\phi$ immediately following a compression step.
We find that the contact percolation transition at $\phi_P = 0.549 \ll
\phi_J$ becomes sharper with system size and obeys finite-size
scaling, but with a correlation length exponent $\nu \approx 1.68$
that is significantly larger than that for random
continuum~\cite{gawlinski} and rigidity percolation~\cite{thorpe}, but
smaller than that found for contact percolation in athermal
particulate systems with short-range attractions~\cite{lois}. (Note
that percolation onset occurs at a similar value $\phi_P = 0.558 \pm
0.008$ for athermal systems with short-range attractions.)  In
contrast, the exponent $\tau \approx 2.01$ that characterizes the
power-law scaling of the cluster size distribution and the fractal
dimension $D \approx 1.89$ are similar to that for random continuum
percolation and contact percolation for athermal systems with
short-range attractions, and obey hyperscaling $D(\tau-1)=2$. (See
Table~\ref{perctable}.)  We find that these results are insensitive to
the damping coefficient ${\widetilde b}$ in the overdamped limit and
compression step for $\Delta \phi \le 10^{-3}$.

\begin{figure}
\includegraphics[width=1.05\columnwidth]{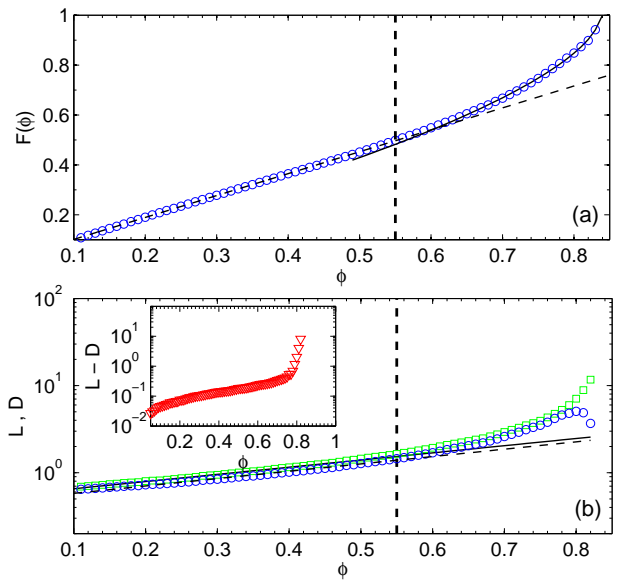}
\vspace{-0.1in}
\caption{(a) The fraction $F(\phi)$ of non-floppy eigenmodes of the
dynamical matrix in the system measured immediately after a
compression step of $\Delta \phi=10^{-3}$ over a range of packing
fractions for $N=1000$.  Fits to the low and high $\phi$ behavior of
$F(\phi)$ are shown as dashed and solid lines, respectively. (b) The
accumulated distance $L$ and displacement $D$ between successive
compressions normalized by $\Delta \phi$ as a function of packing
fraction.  The low-$\phi$, roughly exponential behavior of $L$ and $D$
is indicated by solid and dashed lines.  The inset shows $L-D$ versus $\phi$.
In both panels, the vertical line indicates the percolation transition
$\phi_P = 0.549$.}
\vspace{-0.1in}
\label{fig3}
\end{figure}

We have shown that immediately following each compression step, a
system-spanning cluster of interparticle contacts forms at $t=0$ for
$\phi \ge \phi_P$, much below $\phi_J$.  To determine if this purely
geometrical transition influences the mechanical properties of the
system, we measure the eigenvalues of the dynamical matrix at $t=0$
following each compression step.  As static packings are compressed,
they progressively become less floppy, {\it i.e.}  fewer single and
collective particle motions cost zero energy.  We quantify the
increase in rigidity by measuring the fraction of non-floppy or
`blocked' eigenmodes $F(\phi)$---the ratio of the number $N_{nf}$ of non-zero
eigenvalues of the dynamical matrix to the total number of nontrivial
modes $2N' - 2$ (where $N'=N-N_r$ and $N_r$ is the number of rattler
particles at jamming~\cite{footnote})---over a range of packing
fractions. With this definition, $F(\phi_J)=1$. In Fig.~\ref{fig3}
(a), we show that the fraction of non-floppy modes $F(\phi)$ grows
linearly with $\phi$ for small $\phi$ and near jamming $F(\phi)$
scales as $1 - A (\phi_J - \phi)^{\alpha}$ with $A \approx 0.99$ and
$\alpha \approx 0.6$.  We find that $F(\phi)$ begins to deviate from
linear behavior near $\phi_P$.  In addition, near $\phi_P$, $F(\phi)$
does not obey the scaling behavior near $\phi_J$.  Thus, we have
identified a signature in the mechanical response
near $\phi_P$.

\begin{table}
\begin{center}
\begin{tabular}{lccc}
Percolation type & $\nu$ & $\tau$ & $D$\\ \hline
Repulsive contact & $1.70 \pm 0.09$ & $2.01 \pm 0.04$  & $1.89 \pm 0.03$\\
Attractive contact & $1.92 \pm 0.03$ & $2.04 \pm 0.04$  & $1.88 \pm 0.04$\\
Continuum & $1.34 \pm 0.02$ & $2.02 \pm 0.03$ & $1.91 \pm 0.04$\\
\hline \hline
\end{tabular}
\caption{Critical exponents for contact percolation in athermal
systems with purely repulsive interactions and short-range attractive
interactions~\cite{lois}, as well as random continuum
percolation~\cite{gawlinski} in 2D.  }
\label{perctable}
\end{center}
\vspace{-0.25 in}
\end{table}

In addition, we find that the contact percolation transition at
$\phi_P$ signals the onset of complex spatiotemporal response to
isotropic compression.  We measure two quantities that characterize
cooperative particle motion: 1) the displacement in
configuration space between systems at successive compressions,
\begin{equation}
\label{D}  
D = \sqrt{\sum_{i=1}^N [(x_i(0) - x_i(\infty))^2 + (y_i(0) - y_i(\infty))^2]}, 
\end{equation} 
where $(x_i(0),y_i(0))$ and $(x_i(\infty),y_i(\infty))$ are the
locations of particle $i$ at $t=0$ and $t\rightarrow \infty$,
respectively, during the energy relaxation, and 2) the accumulated
distance traveled in configuration space,
\begin{equation}
L = \int_0^{\infty} dt \sqrt{\sum_{i=1}^N {\vec v}^2_i(t)} 
\end{equation}
from $t=0$ after the compression to the end of the energy relaxation.  Both 
$D$ and $L$ are normalized by the compression step size $\Delta \phi$. 
In Fig.~\ref{fig3} (b), we show that $D$ and $L$ grow roughly
exponentially for small $\phi$, but begin to deviate from the
low-$\phi$ behavior near $\phi_P$.  An even larger signature of
collective, non-affine motion is shown in the inset to Fig.~\ref{fig3}
(b).  The difference $L-D$ begins to increase dramatically at $\phi_l
\approx 0.78$, which is also significantly below $\phi_J$.  Thus, for
$\phi>\phi_P$ particles move collectively in response to compression, but
for $\phi > \phi_l$ after relaxation they end up close to where they started.

\begin{figure}
\includegraphics[width=1.05\columnwidth]{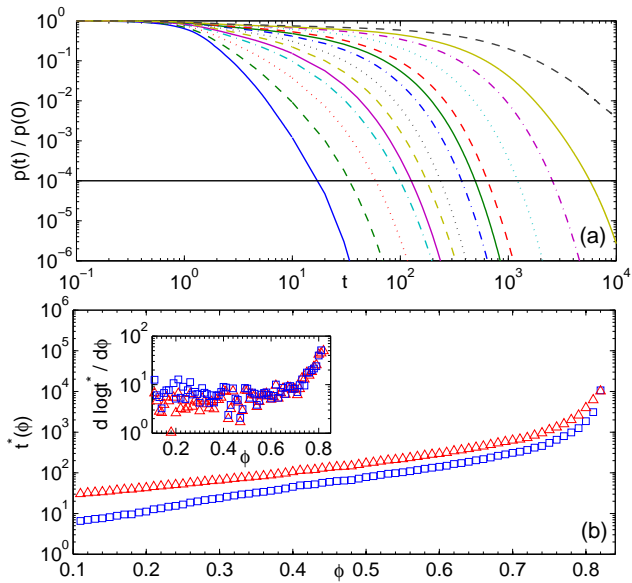}
\vspace{-0.1in}
\caption{ (a) Normalized pressure ${\widetilde p} \equiv p(t)/p(0)$
versus $t$ following a compression step $\Delta \phi=10^{-3}$ with
${\widetilde b}=1$ over a range of packing fractions including
$\phi=0.2$, $0.3$, $0.4$, $0.5$, $0.55$, $0.6$, $0.65$, $0.7$, $0.73$,
$0.75$, $0.78$, $0.8$, $0.81$, and $0.82$ (from left to right)
for $N=200$. The horizontal line ${\widetilde p} = 10^{-4}$
indicates the normalized pressure at which the pressure relaxation
time $t^*$ was measured. (b) The pressure relaxation timescale
$t^*(\phi)$ for systems with ${\widetilde b} \rightarrow \infty$
(triangles) and $1$ (squares).  The inset shows the
logarithmic derivative of $t^*$ with respect to $\phi$. }
\vspace{-0.1in}
\label{fig4}
\end{figure}

We also characterize the time-dependent response to compression.  In
Fig.~\ref{fig4} (a), we plot the normalized pressure ${\widetilde p}
\equiv p(t)/p(0)$ as a function of time following a compression step
$\Delta \phi = 10^{-3}$ over a wide range of packing fractions.  The
normalized pressure develops a plateau near ${\widetilde p}= 0.6$ and
decays more slowly with increasing packing fraction. The plateau in
the normalized pressure suggests that significant decay of the
pressure in the system requires first the collective motion of
exterior particles in percolating clusters followed by motion of the
interior particles.  In Fig.~\ref{fig4} (b), we plot the pressure
relaxation time $t^*$, {\it i.e.} the time at which the normalized
pressure has decayed to a specified small value ${\widetilde p} =
10^{-4}$. The shape of the increase in the pressure relaxation time is
relatively insensitive to ${\widetilde b}$ (for overdamped dynamics)
and ${\widetilde p}$.  In the inset to Fig.~\ref{fig4} (b), the
logarithmic slope of $t^*$ with respect to $\phi$ shows a subtle
increase near $\phi_P$, but a much more dramatic increase near
$\phi_l$, where the the measure of cooperative motion increased
significantly. (See the inset to Fig.~\ref{fig3} (b).) Thus growth in
the stress relaxation time occurs at packing fractions significantly
below $\phi_J$.

In this letter, we described extensive computational studies of the
geometrical and mechanical properties of unjammed, athermal
particulate systems undergoing quasistatic isotropic compression.  A
novel aspect of our work is that we focused on unjammed rather than
jammed configurations~\cite{ostojic}.  We first showed that these
systems undergo the `contact percolation' transition at $\phi_P <
\phi_J$, much below the jamming transition.  Above $\phi_P$, these
systems possess a system-spanning cluster of non-force bearing
interparticle contacts. Near the transition, the cluster size
distribution displays robust scaling with system size with a
correlation length exponent $\nu$ that is larger~\cite{lois} than that
for random continuum and rigidity percolation, and thus contact
percolation belongs to a distinct universality class.  We also
performed measurements to assess the mechanical properties and
time-dependent stress response for unjammed systems with $\phi_P <
\phi < \phi_J$.  We found three key results: 1) the increase in the
fraction $F$ of non-floppy modes begins to deviate from the low-$\phi$
behavior near $\phi_P$ and the acceleration of non-floppy modes near
$\phi_P$ is distinct from the approach to $F=1$ near $\phi_J$. 2) The
$\phi$-dependence of two measures of cooperative motion---the
accumulated distance $L$ and displacement $D$ in configuration space
between successive compressions---deviates from the low-$\phi$
behavior near $\phi_P$.  Further, their difference $L-D$ begins to
grow dramatically near $\phi_l \approx 0.78$, which signals blocking
in configuration space well below $\phi_J$. 3) The growth in the
stress relaxation time $t^*$ also begins to deviate from the
low-$\phi$ behavior near $\phi_l$.  Thus, the process of
rigidification and the onset of cooperative, heterogeneous, and slow
dynamics occurs well below jamming onset in athermal particulate systems.

{\bf Acknowledgments} This research was supported by the National
Science Foundation under Grant Nos. DMS-0835742 (CO, TS), PHY-1019147
(TS), and CBET-0968013 (MS).


\begin{references}
\bibitem{footnote}
Jammed systems possess all positive nontrivial eigenvalues of the dynamical 
matrix except for $d$ zero eigenvalues per rattler particle with 
fewer than $d+1$ contacts, where $d$ is the spatial dimension.
\end{references}
\end{document}